\documentclass[pra,showpacs,twocolumn,showkeys]{revtex4-1}
\usepackage{graphics}
\usepackage{graphicx}
\usepackage{amsmath}
\newcommand{\bibs}{/Users/jperezmo/Dropbox/References/BibFile}

\begin{document}

\title{Three-state interactions determine the second-order nonlinear optical response}

\author{Javier Perez-Moreno}
\email{jperezmo@skidmore.edu} \affiliation{Department of Physics, Skidmore College, Saratoga Springs, New York 12866}

\affiliation{Department of Physics and Astronomy, Washington State University, Pullman, Washington 99164-2814}

\begin{abstract}
Using the sum-rules, the sum-over-states expression for the diagonal term of first hyperpolarizability can be expressed as the sum of three-state interaction terms. We study the behavior of a generic three-state term to show that is possible to tune the contribution of resonant terms by tuning the spectrum of the molecule. When extrapolated to the off-resonance regime, the three-state interaction terms are shown to behave in a similar manner as the three-level model used to derive the fundamental limits. We finally show that most results derived using the three-level ansatz are general, and apply to molecules where more than three levels contribute to the second-order nonlinear response or/and far from optimization.
\end{abstract}

\pacs{42.65.An, 33.15.Kr, 11.55.Hx, 32.70.Cs}

\maketitle

\section{Introduction}

Materials with tailored second-order nonlinear optical properties are needed for the adavance of diverse applications such as information technology,\cite{dalto10.01} bio-imaging,\cite{helmchen2005deep,de12.01,lopez2015thiophene} and cancer therapy.\cite{brown03.01,brown2003dynamic}. The nonlinear optical response in organic materials is originated at the molecular level, and the response of the bulk is related to the molecular response by simple addition rules. This means that organic materials can achieve the fastest response and also that there is an immense pool of potential organic structures suitable for synthesis (of the order of Avogadro's number).\cite{lipinski2004navigating,ertl2003cheminformatics} Which such a huge number of potential structures, a better understanding of the underlying mechanism behind the molecular response is needed in order to tailor organic materials to their full potential.

Applying the {\em ``three level ansatz''} (described below) one can show that the strength of the molecular nonlinear optical response is limited by the number of electrons. When these electrons are optimally arranged the fundamental limit is reached.\cite{kuzyk00.01,kuzyk00.02,kuzyk03.02,kuzyk03.01} The quantum limits analysis has been used to determine molecular efficiency, highligh the mechanisms that determine the molecular response,\cite{tripa04.01, tripa06.01, perez07.02, zhou08.01, perez05.01, perez11.02, perez11.01, perez09.02, van2012dispersion, de12.01} introduce new paradigms for optimization,\cite{perez09.01, perez07.01, perez06.01, perez11.02, Kang05.01, brown08.01, He11.01}, establish fundamental scaling laws,\cite{kuzyk10.01, kuzyk13.01}, and to identify the best molecular candidates for second-order nonlinear applications.\cite{perez16.01} In this paper we generalize these results by deriving expressions that apply to all molecules that can be represented by a second-order nonlinear susceptibility, even when the three-level ansatz does not apply.

\section{Theory}

The property that quantifies the strength of a molecule's second-order nonlinear optical interaction is the first hyperpolarizability, $\beta_{ijk}(-\omega_{1}; \omega_{1},\omega_{2})$, a third-rank tensor that depends on two input frequencies $\omega_{1}$ and $\omega_{2}$, with $\omega_{1}+\omega_{2}=\omega_{\sigma}$. A sum-over-states expression for the first hyperpolarizability can be obtained using time-dependent perturbation theory and the Bogoliubov and Mitrolplsky method of averages and was first derived by Orr and Ward:\cite{orr71.01}
\begin{eqnarray}
\label{eq:betaten}
\beta_{ijk}(-\omega_{\sigma}; \omega_{1},\omega_{2})=-(\hbar)^{-2}
e^{3} I_{1,2} {\sum_{m,n}}' \left\{ \frac{\langle r_{i} \rangle_{0m}
\langle \bar{r}_{k} \rangle_{mn} \langle r_{j}
\rangle_{n0}}{(\Omega_{m0}-\omega_{\sigma})
(\Omega_{n0}-\omega_{1})} \right. \nonumber \\ + \left. \frac{
\langle r_{k} \rangle_{0m} \langle \bar{r}_{j} \rangle_{mn} \langle
r_{i} \rangle_{n0}}{(\Omega^{*}_{m0}+\omega_{2})
(\Omega^{*}_{n0}-\omega_{\sigma})} +\frac{ \langle r_{k}
\rangle_{0m} \langle \bar{r}_{i} \rangle_{mn} \langle r_{j}
\rangle_{n0}}{(\Omega^{*}_{m0}+\omega_{2})
(\Omega_{n0}-\omega_{1})}\right\},
\end{eqnarray}
where the prime in the sum indicates that the ground state is excluded from the sum ($n \neq 0$ and $m \neq 0$), $\hbar$ is the reduced Plank constant and $(-e)$ is the charge of an electron. The $ith$ component of the position operator is represented by $r_{i}$, with matrix elements (between states $| m \rangle$ and $\ n \rangle$):
\begin{equation}
\langle r_{i} \rangle_{mn} = \langle m | r_{i} | n \rangle,
\end{equation}
and the operator $I_{1,2}$ averages over all terms generated by pairwise permutations of $(i,\omega_{1})$ and $(j,\omega_{2})$ in the expression. Notice that we have used the notation $\Omega_{n0} = \Omega_{n} - \Omega_{0}$, and that the barred operator is defined as:
\begin{equation}
\bar{O}=O-\langle O \rangle_{00}.
\end{equation}

By assumption, the eigenvalues of the time-independent Hamiltonian, $H(0)$ (which describes the isolated molecule before the interaction with light is turned on), are complex to allow for natural decay:
\begin{equation}
H(0)|n\rangle = \hbar \Omega_{n} | n \rangle = \hbar \left( \omega_{n} - i \frac{\Gamma_{n}}{2}\right) |n \rangle, 
\end{equation}
such as the energy of the eigenstate $|n \rangle$ is given by $\hbar \omega_{n} = E_{n}$ and its inverse radiative lifetime is $\hbar \Gamma_{n}$.

For molecules that are approximately 1-dimensional (that is, for molecules where the conjugated path has $C_{\infty v}$ symmetry) the diagonal term of the first hyperpolarizability dominates over all other components. Choosing the geometry such as the $x$-axis coincides with the axis of the molecule, the diagonal component of the first hyperpolarizability can be expressed as:
\begin{equation}
\label{eq:betadiag}
\beta_{xxx}(-\omega_{\sigma};\omega_{1},\omega_{2})=(-e)^{3} {\sum_{mn}}' x_{0m} \bar{x}_{mn} x_{n0} \cdot D^{(2)}_{mn}(\omega_{1};\omega_{2}),
\end{equation}
where $x_{mn} = \langle m | r_{x} | n \rangle$ and the dispersion factors are defined as:
\begin{widetext}
\begin{eqnarray}
D^{(2)}_{mn}(\omega_{1},\omega_{2}) = \frac{1}{\hbar^{2}} I_{1,2} \left\{ \frac{1}{(\Omega_{m0}-\omega_{\sigma})
(\Omega_{n0}-\omega_{1})} \right. &+& \left. \frac{1}{(\Omega^{*}_{m0}+\omega_{2})
(\Omega^{*}_{n0}+\omega_{\sigma})} +\frac{1}{(\Omega^{*}_{m0}+\omega_{2})
(\Omega_{n0}-\omega_{1})} \right\} \nonumber \\
= \frac{1}{2 \hbar^{2}} \left\{ \frac{1}{(\Omega_{m0}-\omega_{\sigma})
(\Omega_{n0}-\omega_{1})} \right. &+& \left. \frac{1}{(\Omega^{*}_{m0}+\omega_{2})
(\Omega^{*}_{n0}+\omega_{\sigma})} +\frac{1}{(\Omega^{*}_{m0}+\omega_{2})
(\Omega_{n0}-\omega_{1})} \right. \nonumber \\
+ \left. \frac{1}{(\Omega_{m0}-\omega_{\sigma})
(\Omega_{n0}-\omega_{2})} \right. &+& \left. \frac{1}{(\Omega^{*}_{m0}+\omega_{1})
(\Omega^{*}_{n0}+\omega_{\sigma})} +\frac{1}{(\Omega^{*}_{m0}+\omega_{1})
(\Omega_{n0}-\omega_{2})} \right\},
\label{eq:betadis}
\end{eqnarray}
\end{widetext}
where, for clarity, we have explicitly performed the averaging indicated by $I_{1,2}$.

Far away from resonances, $E_{m0} \gg \hbar \omega_{1}$, $E_{m0} \gg \hbar \omega_{2}$ and $E_{m0} \gg \frac{\hbar \Gamma_{m}}{2}$, so we can approximate Eq. \ref{eq:betadis} as:
\begin{equation}
D_{mn}^{(2)} \approx \frac{3}{E_{mg}E_{n0}},
\end{equation} 
such as that, in the off-resonance regime, Eq. \ref{eq:betadiag} is approximated by:
\begin{equation}
\label{eq:betaoff}
\beta_{xxx}^{\mbox{off}} \approx (-e)^{3} {\sum_{mn}}' \frac{x_{0m} \bar{x}_{mn} x_{n0}}{E_{m0}E_{n0}}.
\end{equation}

The sum-over-states expressions for the first hyperpolarizability (Eqs. \ref{eq:betaten}, \ref{eq:betadiag} and \ref{eq:betaoff}) depend on an infinite set of transition dipole moments and energies. However, these parameters are not independent. They are related to each other through the Thomas-Kuhn sum rules which apply quite generally to most quantum systems. The first sum rules were originally derived by Thomas and Kuhn using a semiclassical approach.\cite{thom25.01, kuhn25.01} Heisenberg derived them using quantum mechanics principles,\cite{heise25.01} and they were generalized by Bethe et al.\cite{bethe77.01} Here we use the generalized sum rules as derived by Kuzyk.\cite{kuzyk00.01,kuzyk00.02,kuzyk03.02,kuzyk03.01} Choosing the geometry such as the $x$-axis coincides with the axis of the molecule,  $r_{i}=r_{x}$, and denoting $x=\langle r_{x} \rangle$, the sum rules can be expressed as:
\begin{equation}
\label{eq:diagsumrules}
\sum_{n} \left( 2E_{n0} - E_{k0} - E_{l0} \right) x_{kn} x_{nl} = \frac{\hbar^{2} N}{m}\delta_{kl},
\end{equation}
where $N$ is the number of effective electrons in the system, and $\delta_{kl}$ is the Kronecker delta. 

It is important to notice that Eq. \ref{eq:diagsumrules} is actually an infinite set of equations, depending on which specific values of $(k,l)$ are picked. However, since by definition of the inner product $x_{kn} = x_{nk}^{*}$, the sum rule that we obtain by picking $k=s$ and $l=t$ is the complex conjugate of the sum rule that we would obtain by picking $k=t$ and $l=s$. This means that if $k=l$, the corresponding sum rule is real. Also, in Eq. \ref{eq:diagsumrules}, the sum over the dummy index $n$ does include the ground state, while in the sum-over-state expressions (Eqs. \ref{eq:betaten}, \ref{eq:betadiag} and \ref{eq:betaoff}) the ground state is excluded from the sum.

The fundamental limit is obtained by applying the three-level ansatz which assumes that a three-level model accurately describes any quantum system whose nonlinear-optical response is close to the fundamental limit. In other words, the Three-Level Ansatz can be stated as:\cite{shafe13.01}
\begin{quote}
{\em ``When the hyperpolarizability of a quantum system is at its fundamental limit, only three states contribute to the response.''} 
\end{quote} 

Applying the three-level ansatz to Eqs. \ref{eq:diagsumrules} and \ref{eq:betaoff} one can show that the first hyperpolarizability is bounded by the fundamental limit:\cite{kuzyk00.01,kuzyk03.02} 
\begin{equation}
\beta_{xxx}^{\mbox{off}} \leq \beta_{xxx}^{\mbox{max}} = \sqrt[4]{3} \left( \frac{e \hbar}{\sqrt{m}} \right)^{3} \frac{N^{3/2}}{E_{10}^{7/2}},
\label{eq:fundlim}
\end{equation}
and that the expression of the off-resonant first hyperpolarizability (Eq. \ref{eq:betaoff}) simplifies to:
\begin{equation}
\beta^{\mbox{off}}_{xxx}(E,X) = \beta_{xxx}^{\mbox{max}} \cdot f(E) \cdot G(X),
\label{eq:betaEX}
\end{equation}
where the functions $f(E)$ and $G(X)$ are defined as:
\begin{equation}
f(E)=\frac{1}{2} (1-E)^{3/2} (2+3E+2E^{2})
\label{eq:fE}
\end{equation}
and
\begin{equation}
G(X)=\sqrt[4]{3} X \sqrt{\frac{3}{2}(1-X^{4})}.
\label{eq:GX}
\end{equation}
The dimensionless parameters $E$ and $X$ defined as:
\begin{equation}
E=\frac{E_{10}}{E_{20}},
\label{eq:Edef}
\end{equation}
and
\begin{equation}
X= \frac{|x_{01}|}{\sqrt{\frac{\hbar^{2} N}{2 m E_{10}}}}.
\label{eq:Xdef}
\end{equation}

The three-level ansatz has not been rigorously proven but the calculation of the quantum limits is consistent with experimental data and with numerical studies. Experimentally, the first hyperpolarizability of any molecule has always been found to be below the quantum limit.\cite{tripa04.01, tripa06.01, perez07.01, perez09.01} Numerical studies also show that the best quantum systems have maximum hyperpolarizabilities that do not surpass the quantum limit.\cite{zhou06.01, zhou07.02, watki11.01, watki09.01, kuzyk06.02}  

\subsection{Dipole-free expression for the first hyperpolarizability}

The sum-over-states expressions for the first hyperpolarizability (Eqs. \ref{eq:betaten}, \ref{eq:betadiag} and \ref{eq:betaoff}) have been extensively used to model the second-order nonlinear response and to analyze experimental data since they were introduced in 1971.\cite{orr71.01} However, the sum-over-states expressions treat the set of $\{E_{n},x_{mn}\}$ as independent parameters and we know that the sum rules impose constraints over the set. Thus, the traditional sum-over-states expressions are over-specified and require redundant information in order to be evaluated. Furthermore, results based on the optimization of the sum-over-states expressions will treat the parameters as independent, which can lead to erroneous conclusions. 

A more compact sum-over-states expressions that eliminates some of the redundant information by incorporating the sum-rules was introduced by Kuzyk.\cite{kuzyk05.01a} The expression is called ``dipole-free'' since it eliminates the explicit dependence on dipolar terms (ie. terms that require a change in dipole moment). We notice that by definition:
\begin{equation}
\bar{x}_{mn} = 
\begin{cases}
    x_{mn},& \text{if } m \neq m\\
    x_{nn}-x_{00} = \Delta x_{n0}, & \text{if } m = n 
\end{cases}
\end{equation} 
such as the traditional sum-over-states expression for the diagonal term of the first hyperpoloarizability (Eq.\ref{eq:betadiag}) can be expressed as:
\begin{eqnarray}
\beta_{xxx}(-\omega_{\sigma};\omega_{1},\omega_{2})=(-e)^{3} \left( {\sum_{n}}' |x_{n0}|^{2} \Delta x_{n0} \cdot D^{(2)}_{nn}(\omega_{1};\omega_{2}) \right. \nonumber \\
+ \left. {\sum_{n}}' {\sum_{m\neq n}}' x_{0n}x_{nm}x_{m0} \cdot D^{(2)}_{mn}(\omega_{1};\omega_{2}) \right).
\label{eq:betasplit}
\end{eqnarray}
The first sum is made up from all the terms that explicitly require a change in dipole moment ($\Delta x_{n0}=x_{nn}-x_{00}$) in order to contribute. These terms depend only on the transition moments of two states (ground and $n$), while the terms in the second sum connect transition moments of three different states (ground, $n$ and $m$ with $n \neq m$). 

To derive the dipole-free expressions we must consider the sum rules that we obtain by picking $l=0$ and $k \neq 0$ in Eq. \ref{eq:diagsumrules}, and multiplying the resulting expression by $x_{0k}$:
\begin{equation}
\sum_{\substack{n \neq 0 \\ n \neq k}} (2E_{n0}-E_{k0}) x_{0k}x_{kn}x_{n0} + E_{k0}|x_{k0}|^{2} \Delta x_{k0} =0,
\end{equation}
Rearranging terms we arrive to:
\begin{equation}
|x_{0n}|^{2} \Delta x_{n0} = - \sum_{{\substack{n \neq 0 \\ n \neq k}}} \frac{2E_{m0}-E_{n0}}{E_{n0}} x_{0n}x_{nm}x_{m0}.
\label{eq:dipolechange}
\end{equation}
where by assumption $n \neq 0$. Substituting Eq. \ref{eq:dipolechange} into Eq. \ref{eq:betasplit} leads to the dipole-free expressions for the first hyperpolarizability:\cite{kuzyk05.01a}
\begin{equation}
\beta_{xxx}(-\omega_{\sigma};\omega_{1},\omega_{2})=(-e)^{3} {\sum_{\substack{n \neq 0 \\ n \neq k}}}' x_{0n}x_{nm}x_{m0} \cdot H_{nm}^{(2)},
\label{eq:dipolefree}
\end{equation} 
were we have defined the energy terms:
\begin{equation}
H_{nm} = \left( D^{(2)}_{mn}(\omega_{1};\omega_{2}) -\frac{(2E_{m0}-E_{n0})}{E_{n0}} \cdot D^{(2)}_{nn}(\omega_{1};\omega_{2}) \right).
\label{eq:Hterms}
\end{equation}
For simplicity of notation we have omitted the dependence on the input frequencies in the definition of $H_{nm}$.

It is important to notice that the only new assumption that has been made in the derivation of Eq. \ref{eq:dipolefree} from Eq. \ref{eq:betadiag} is that the $(k,0)$ sum rules with $k \neq 0$ are obeyed. The sum rules have been shown to apply to the most general form of the Hamiltonian for $N$ electrons of mass $m$ that interact through electromagnetic forces. They hold for any scalar potential that is a function of the position of the electrons, spin angular momentum and a linear function of the orbital angular momentum.\cite{kuzyk13.01} Therefore, they apply quite generally to all molecules. Only exotic potentials that are not physically meaningful can lead to violation of the generalized sum rules. Furthermore, while truncation of the sum-rules to the contribution of few states might lead to inaccuracies,\cite{kuzyk14.01} the derivation of the dipole-free expression does not assume truncation of the sum-rules and therefore the expression is exact. 

\section{Results}

The dipole-expression for the first hyperpolarizability is still over-specified in the sense that it does not take into account the relationship between pairs of transition moments:
\begin{equation}
x_{nm} = \langle n | \hat{x} | m \rangle  = \left( \langle m | \hat{x} | n \rangle \right)^{*} = x_{mn}^{*}
\label{eq:trans}
\end{equation}
which follows from the definition of the inner product and the fact that the position operator $\hat{x}$ is real,\cite{griff05.01} and therefore always applies. This implies that the products of transition moments that appear in Eq. \ref{eq:dipolefree} are connected through:
\begin{equation}
x_{0n}x_{nm}x_{m0} = \left( x_{0m}x_{mn}x_{n0} \right)^{*}.
\label{eq:treetrans}
\end{equation} 
Our next goal is to use the relationships between the transition dipole moments to further simplify Eq. \ref{eq:dipolefree}. This is very useful when the transition dipole moments are real as we shall see below, and leads to some general results. 

By explicitly pairing the terms that are connected through Eq. \ref{eq:treetrans} we can rewrite the expression as:
\begin{eqnarray}
\beta_{xxx}(-\omega_{\sigma};\omega_{1},\omega_{2}) = &&\nonumber \\
(-e)^{3} {\sum_{n}}'{\sum_{m > n}}' \left( x_{0m}x_{mn}x_{n0} \cdot H_{mn} + x_{0n} x_{nm} x_{m0} H_{nm} \right) = && \\ \nonumber
(-e)^{3} {\sum_{n}}'{\sum_{m > n}}' \left( x_{0m}x_{mn}x_{n0} \cdot H_{mn} + (x_{0m} x_{mn} x_{n0})^{*} \cdot H_{nm}  \right). &&
\label{eq:betapairs}
\end{eqnarray}
The advantage of Eq. \ref{eq:betapairs} is that now the expression for the first hyperpolarizability is expressed as a sum where each term includes {\em all} the possible contributions of three specific states (ground $|0 \rangle$, $|n \rangle $ and $|m \rangle$). Also, by connecting the conjugated transition moments we have reduced the number of explicit terms that need to be evaluated to compute the first hyperpolarizability by a factor of 2.

More importantly, Eq. \ref{eq:betapairs} clearly highlights how the second-order nonlinear optical response is determined by three-state interactions. In fact, the minimum number of states that must have non-zero transition moments are three. In other words, a strict two-level model (that considers only the contributions of two states) can not lead to second-order nonlinear optical response. However, a typical approximation for the traditional expression of the first hyperpolarizability is taking by assuming that the contribution of two states dominate the response.\cite{oudar77.03} The two-level model has been shown to be unphysical for molecules that can not be approximated as 1-dimensional systems.\cite{bidault2007role, brasselet1996relation, weibel2003quantum} However, according to Eq. \ref{eq:betapairs} (or Eq. \ref{eq:dipolefree}), in order to be consistent with the sum rules, at least three levels must contribute to the response, even for structures that can be approximated to be 1-dimensional.

\subsection{Real transition dipole moments}

We will now make use of a theorem concerning time invariance, as stated by Sakurai:\cite{sakur94.01}
\begin{quote}
{\em ``Suppose the Hamiltonian is invariant under time reversal and the energy eigenstate $|n \rangle$ is nondegenerate; then the corresponding energy eigenfunction is real.''}
\end{quote}

First we notice that in any 1-dimensional system, the solutions to the Schr\"{o}dinger Equation are non-degenerate, so as long as our approximation of treating the system as 1-dimensional holds, this condition is fulfilled. Also, if we assume that the Hamiltonian that describes the molecule is conservative then we can apply the theorem and conclude that the eigenfunctions are real. More generally, we can assume that the eigenfunctions are real when the potential function depends on operators that are time-reversal invariant (such as position, energy, electric field, electric polarization or charge density). The assumption will not apply for potentials that depend on quantities that are not invariant under time-reversal (such as angular momentum or magnetic field). Therefore the following results will apply generally to molecules where relativistic and magnetic effects can be ignored.     

With real eigenfunctions, the transition dipole moments have to be real, which implies that $x_{0n}x_{nm}x_{m0} = x_{0m}x_{mn}x_{n0}$, such as Eq. \ref{eq:betapairs} is simplified to:
\begin{equation}
\beta_{xxx}(-\omega_{\sigma};\omega_{1},\omega_{2}) = (-e)^{3} {\sum_{n}}'{\sum_{m > n}}' x_{0n} x_{nm} x_{m0} \cdot F_{nm},
\label{eq:betarealtrans}
\end{equation}
with:
\begin{equation}
F_{nm}=H_{mn}+H_{nm}.
\label{eq:fnm}
\end{equation}
Thus, if the transition dipole moments are real each term in the sum can be written as the product of a function that explicitly depends on transition dipole moments and a function that explicitly depends on energies. Using Equation \ref{eq:Hterms}, the energy functions $F_{nm}$ can be expressed as:
\begin{eqnarray}
F_{nm}= \left( D^{(2)}_{mn}(\omega_{1};\omega_{2}) -\frac{(2E_{m0}-E_{n0})}{E_{n0}} \cdot D^{(2)}_{nn}(\omega_{1};\omega_{2}) \right. \\ \nonumber
\left. +D^{(2)}_{nm}(\omega_{1};\omega_{2}) -\frac{(2E_{n0}-E_{m0})}{E_{m0}} \cdot D^{(2)}_{mm}(\omega_{1};\omega_{2}) \right).
\label{eq:Fnm}
\end{eqnarray} 
We notice that the contributions from the dispersion terms $D^{(2)}_{nn}$ and $D^{(2)}_{mm}$ are ``weighted'' by functions that depend on energy ratios.  The contribution from $D^{(2)}_{nn}$ is weighted by the factor:
\begin{equation}
f^{weight}_{nn}=-\frac{(2E_{m0}-E_{n0})}{E_{n0}}, 
\end{equation} 
which approaches its maximum value $-1$ if the two energies are very close to each other; and decreases without bound as the difference of energies increases. The contribution from $D^{(2)}_{mm}$ is weighted by the factor:
\begin{equation}
f^{weight}_{mm}-\frac{(2E_{n0}-E_{m0})}{E_{m0}},
\end{equation}
which approaches its minimum value $-1$ when the two energies are very close; reaches zero when the difference of energies is such as $2E_{n0}=E_{m0}$; and approaches its maximum value $1$ when the difference of energies becomes very large. This suggest that it is possible to selectively tune the contribution of resonant terms by targeting molecules with a specific spectrum.

\subsection{Resonant response}

\begin{figure}
  \centering
  \includegraphics[width=3.5in]{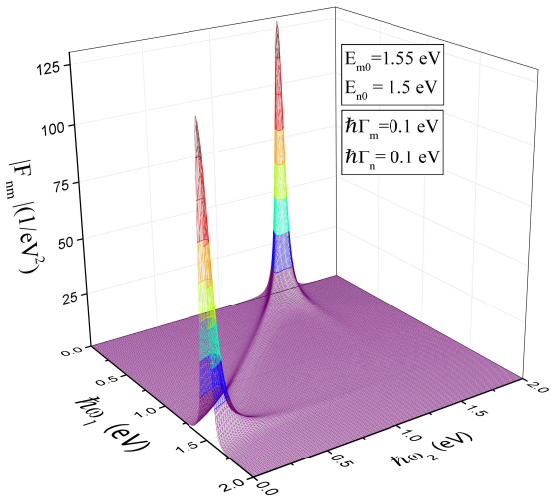}\\
  \caption{Plot of the absolute value of $F_{nm}$ (Eq. \ref{eq:Fnm}) as a function of the photon energies, with $E_{m0}=1.55$ eV, $E_{n0}=1.5$ eV and linewidths set to $0.1$ eV.} \label{fig:FnmDegenerate01}
\end{figure}

\begin{figure}
  \centering
  \includegraphics[width=3.5in]{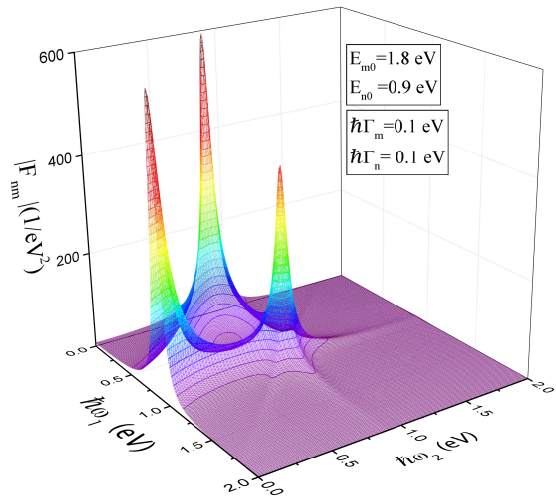}\\
  \caption{Plot of the absolute value of $F_{nm}$ (Eq. \ref{eq:Fnm}) as a function of the photon energies, with $E_{m0}=1.8$ eV, $E_{n0}=0.9$ eV and linewidths set to $0.1$ eV.} \label{fig:FnmEnHalf}
\end{figure}

\begin{figure}
  \centering
  \includegraphics[width=3.5in]{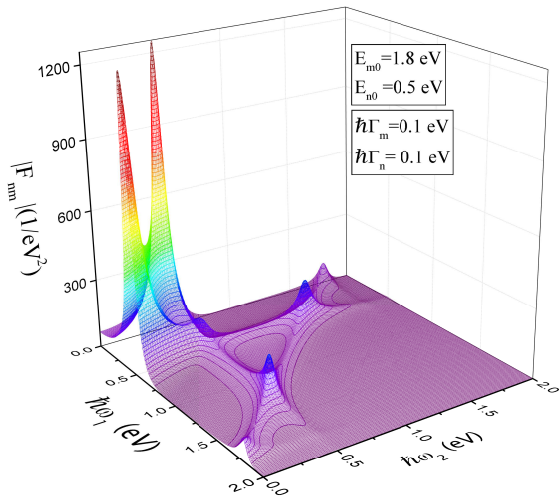}\\
  \caption{Plot of the absolute value of $F_{nm}$ (Eq. \ref{eq:Fnm}) as a function of the photon energies, with $E_{m0}=1.8$ eV, $E_{n0}=0.5$ eV and linewidths set to $0.1$ eV.} \label{fig:FnmEnFar}
\end{figure}

To determine how the weighting factors affect the overall resonant response we will consider some significant combinations of energies. Typical values for the energy differences on a molecule range from 1 to 3 eV, with linewitdhs ($\hbar \Gamma_{n}$) ranging between 0.1 and 0.5 eV. Let us consider first the effect of the energy spectrum by setting all the linewidths to $\hbar \Gamma_{n}=\hbar \Gamma_{m} = 0.1$ eV, and three significant energy distributions as shown in Figures \ref{fig:FnmDegenerate01}, \ref{fig:FnmEnHalf} and \ref{fig:FnmEnFar}.

Figure \ref{fig:FnmDegenerate01} plots the absolute value of $F_{nm}$ as a function of the photon energies with $E_{m0}=1.55$ eV and $E_{n0}=1.5$ eV. The weighting factors are $f^{weight}_{nn}=-1.07$ and $f^{weight}_{mm}=-0.9$. The linewidths are set to $0.1$ eV. As expected, the resonances occur when the combination $\hbar (\omega_{1}+\omega_{2})$ matches one of the energy values. In this case, the energy values are very close such as the two resonances add up and the net result looks like a single resonance. The highest resonant values are achieved when one of the photon energies approaches $0$ eV, where $|F_{nm}|$ peaks and reaches its maximum value ($\approx 125$ eV$^{-2}$). If the resonance is such that none of the photon energies is close to zero, $|F_{nm}|$ decreases significantly (about an order of magnitude). Finally, if the photon energies miss the resonance by more than 0.3 eV, $|F_{nm}|$ becomes pretty flat and approaches zero.

Figure \ref{fig:FnmEnHalf} plots the absolute value of $F_{nm}$ as a function of the photon energies with $E_{m0}=1.8$ eV and $E_{n0}=0.9$ eV. The weighting factors become $f^{weight}_{nn}=-3.0$ and $f^{weight}_{mm}=0$. The linewidths are set to $0.1$ eV. We can distinguish three mean resonances occurring when  $\hbar (\omega_{1}+\omega_{2}) \approx E_{n0}$, $\hbar \omega_{1} \approx E_{n0}$ or $\hbar \omega_{2} \approx E_{n0}$. Interestingly, we do not see any resonances when the photon energies match $E_{m0}$ which must be due to the fact that the dispersion term $D^{(2)}_{mm}(\omega_{1};\omega_{2})$ does not contribute to the energy function. Overall, the effects of the resonances have been scaled by a factor of 5 when compared with Figure \ref{fig:FnmDegenerate01}. Again, the peaks occur when one of the photon approaches $0$ eV and the other matches $E_{n0}$. This is the regime of operation for the electro-optic effect. There is also another significant peak when each photon energy matches $E_{n0}$, which correspond to the regime of operation of second-harmonic generation.
 
Figure \ref{fig:FnmEnFar} plots the absolute value of $F_{nm}$ as a function of the photon energies with $E_{m0}=1.8$ eV and $E_{n0}=0.5$ eV. The weighting factors are $f^{weight}_{nn}=-5.2$ and $f^{weight}_{mm}=0.35$. The linewidths are set to $0.1$ eV. Now we can see resonances when the photon energies match both $E_{n0}$ and $E_{m0}$, but the resonance effects due to $E_{n0}$ are enhanced. As before, the maxima occur when one of the photon energies approaches $0$ eV and the other matches $E_{n0}$ (electro-optic regime). 

Further exploration confirms that trends shown in Figures \ref{fig:FnmDegenerate01}, \ref{fig:FnmEnHalf} and \ref{fig:FnmEnFar} are general and do not depend on the specific values of $E_{m0}$ and $E_{n0}$. The resonant effects are modulated through the weighting functions. Close to degeneracy, the resonant effects are minimized, which must be due to an overall cancellation effect (quantum interference). As the energy difference increases, the resonant effects are enhanced, specially resonances associated with the smallest energy $E_{n0}$ which is explained by the fact that $f^{weight}_{nn}$ becomes large in magnitude. In all the cases, the best response corresponds to the regime of operation of the electro-optic effect. Interestingly, the best energy spacing for on-resonant second-harmonic generation occurs when the energies are spaced like a two-state quantum harmonic oscillator ($E_{m0}=2 E_{n0}$). This could be used to design more efficient molecules for second-harmonic generation imaging where the effects of resonance can be exploited to achieve spectroscopic selectivity.\cite{de12.01}

\begin{figure}
  \centering
  \includegraphics[width=3.5in]{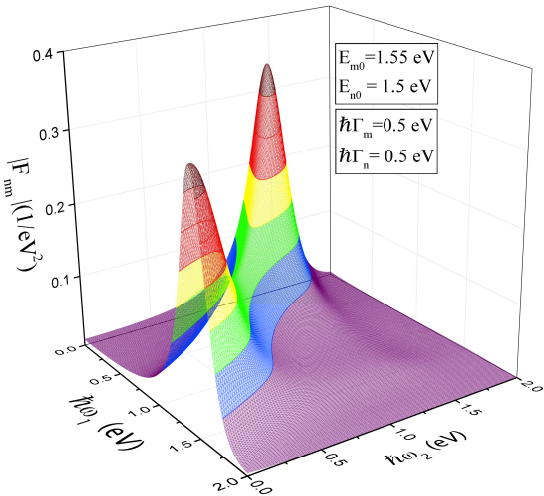}\\
  \caption{Plot of the absolute value of $F_{nm}$ (Eq. \ref{eq:Fnm}) as a function of the photon energies, with $E_{m0}=1.55$ eV, $E_{n0}=1.5$ eV and linewidths set to $0.5$ eV.} \label{fig:FnmDege05}
\end{figure}

In general, as the linewidths broaden, the resonance effects dilute and the absolute value of $|F_{nm}|$ decreases dramatically. Figure \ref{fig:FnmDege05} plots the absolute value of $F_{nm}$ as a function of the photon energies with $E_{m0}=1.55$ eV and $E_{n0}=1.5$ eV, with the linewidths set to $0.5$ eV. If we compare it with Figure \ref{fig:FnmDegenerate01} (with the same energy values) we can see how although the overall shape of the function is similar, the broadening of the linewidths by a factor of 5 results in a decrease of the peak response by 2 orders of magnitude. 

\begin{figure}
  \centering
  \includegraphics[width=3.5in]{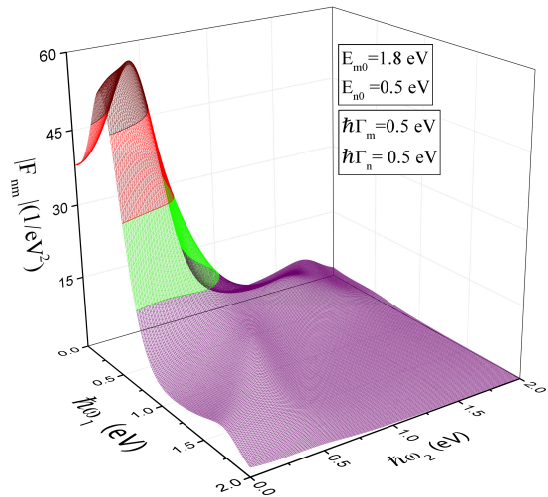}\\
 \caption{Plot of the absolute value of $F_{nm}$ (Eq. \ref{eq:Fnm}) as a function of the photon energies, with $E_{m0}=1.8$ eV, $E_{n0}=0.5$ eV and linewidths set to $0.5$ eV.} \label{fig:FnmFar05}
\end{figure}

Figure \ref{fig:FnmFar05} plots the absolute value of $F_{nm}$ as a function of the photon energies with $E_{m0}=1.8$ eV and $E_{n0}=0.5$ eV, with linewidths set to $0.5$ eV. In comparison with Figure \ref{fig:FnmEnFar} (with the same energy values), the shape of the resonances has been mostly diluted and that the peak values due to the resonances at $E_{n0}$ are 20 times smaller.   

In conclusion, both the ratio of energies and the magnitudes of the linewidths have a strong influence on the shape and magnitude of the energy function $F_{nm}$. By tuning the energy ratio, we can selectively minimize or enhance the resonance effects. With regards to the photon energies, the highest response is always achieved in the electro-optic regime. However, if we want to improve the resonant response for second-harmonic applications, we must design molecules with $E_{m0}=2E_{n0}$.
 
\subsection{Off-resonance response}

Far away from resonances, the dispersion factors $D_{mn}$ are approximated by Equation \ref{eq:betadis}, such as the energy functions become:
\begin{equation}
F_{nm}^{\mbox{off}}=3 \left( \frac{2}{E_{m0}E_{n0}} - \frac{(2E_{m0}-E_{n0})}{E_{n0}^{3}} -\frac{(2E_{n0}-E_{m0})}{E_{m0}^{3}} \right).
\label{eq:fnmoff}
\end{equation}

First, we notice that $F_{nm}^{\mbox{off}}$ diverges when $E_{m0} \rightarrow \infty$, which is due to the divergence of the weighting function $f_{nm}^{weight}$. Experimentally the values of the first hyperpolarizability are clearly bounded. Furthermore, in order for the sum-over-states approach to be valid, the full expression must be convergent. Thus, there must be some other mechanism that prevents this divergent behavior. Taking a hint from the derivation of the quantum limits, the problem can be resolved if the dependence on the transition dipole moments on energies is such that the divergence is canceled.\cite{kuzyk00.01} We can indeed proof this using the remaining set of sum rules.

\subsubsection{Generalized scaling laws}

The derivation of the dipole-free expression (Eq. \ref{eq:dipolefree}) uses the subset of sum-rules that is obtained by picking $l=0$ and $k \neq 0$ in the general expression (Eq. \ref{eq:diagsumrules}). If instead, we pick $l=k=m$ we obtain the following subset:
\begin{equation}
\frac{2m}{\hbar^{2} N} \cdot \sum_{n=0} (E_{n0}-E_{m0}) |x_{nm}|^{2} = 1.
\end{equation}
The expression on the left must remain bounded (and equal to 1) for any combination of energies and in the limiting cases when $E_{n0} \rightarrow E_{m0}$ and $|(E_{n0}-E_{m0})| \rightarrow \infty$. This leads to the following generalized scaling law for the transition dipole moments:
\begin{equation}
\frac{2m}{\hbar^{2}N}|x_{mn}|^{2} \propto \frac{1}{(E_{m0}-E_{n0})},
\end{equation} 
which implies:
\begin{equation}
|x_{mn}| \propto \sqrt{\frac{\hbar^{2} N}{2 m} \frac{1}{(E_{m0}-E_{n0})}}.
\label{eq:genscaling}
\end{equation}
Using Eq. \ref{eq:genscaling} we can determine the energy dependence of a generic term in the sum-over-states as:
\begin{equation}
x_{0m}x_{mn}x_{n0} \cdot F_{nm}^{\mbox{off}} \propto \left( \sqrt{\frac{\hbar^{2} N}{2 m}} \right)^{3} \sqrt{\frac{1}{E_{m0}E_{n0}(E_{m0}-E_{n0})}} \cdot F_{nm}^{\mbox{off}},
\end{equation} 
which, after some manipulation leads to:
\begin{equation}
(-e)^3 x_{0m}x_{mn}x_{n0} \cdot F_{nm}^{\mbox{off}} = k_{nm} \left(e \sqrt{\frac{\hbar^{2} N}{2 m}} \right)^{3} \frac{1}{E_{n0}^{7/2}} \cdot f(\frac{E_{n0}}{E_{m0}}).
\label{eq:gentermf}
\end{equation}
where $k_{nm}$ must be a function of $n$ and $m$ that does not depend explicitly on energies; and $f(E)$ is the same energy function that one obtains using the three-level ansatz, but now applied to the generalized energy ratio $E_{n0}/E_{m0}$. This ratio is bounded between 0 and 1, since by definition $E_{m0}>E_{n0}$. The behavior of $f(E)$ is well known. For all possible ratios of energies $f(E)$ is a well defined monotonically increasing function, that reaches its maximum value at $f(0)=1$ and its minimum value at $f(1)=0$. This is a remarkable result that shows that aside from the scaling factor $E_{n0}^{-7/2}$, the energy dependence of a generic term in the sum-over-states mirrors the energy dependence that is obtained using the three-level ansatz. However, while the three level-ansatz assumes that {\em only} three states contribute to the response, no such assumption has been used to derive Equation \ref{eq:gentermf}.

In fact, if we define the generalized energy function as:
\begin{equation}
f^{\mbox{\small{gen}}}_{nm}= \left( \frac{E_{10}}{E_{n0}} \right)^{7/2} \cdot f(\frac{E_{n0}}{E_{m0}}),
\label{eq:genenfunct}
\end{equation}
we can rewrite the contribution of a generic term in the sum-over-states as:
\begin{equation}
(-e)^3 x_{0m}x_{mn}x_{n0} \cdot F_{nm}^{\mbox{off}} = \beta_{xxx}^{\mbox{max}} \cdot G_{nm}^{\mbox{\small{gen}}} \cdot f_{nm}^{\mbox{\small{gen}}},
\label{eq:gentermkto}
\end{equation} 
where $G_{nm}^{\mbox{\small{gen}}}$ is another function that results from combining the factor $k_{nm}$ with fundamental constants. By construction, $G_{nm}^{\mbox{\small{gen}}}$ is bounded and does not depend explicitly on energies. Also, since by definition $f^{\mbox{\small{gen}}}_{mn}$ is dimensionless, $G_{mn}^{\mbox{\small{gen}}}$ must also be a dimensionless quantity.

In conclusion, using the sum-rules, the expression for $\beta^{\mbox{off}}_{xxx}$ can be written as the sum of three-state interactions (ground and two excited states). The functional behavior of each three-state contribution is the same and can be expressed as the product of a function that depends on the distribution of transition dipole moments, a function that only depends on energy ratios and the fundamental limit. This generalizes the results derived using the three level ansatz (Eqs. \ref{eq:fundlim} to \ref{eq:GX}) but applies to all systems irregardless on how many states contribute to the response. Notice that when the minimum amount of states contribute to the response ($m=2$ and $n=1$),  $f^{\mbox{\small{gen}}}_{mn}$ does automatically become $f(E)$, but $G^{\mbox{\small{gen}}}_{mn}$ does not become $G(X)$ unless further assumptions are made.

\section{Applications}

\subsection{Generalized scaling law for the first hyperpolarizability}

Using Eq. \ref{eq:gentermkto} the sum-over-states expression becomes:
\begin{equation}
\beta^{\mbox{off}}_{xxx}= \beta_{xxx}^{\mbox{max}} \cdot \left( {\sum_{n}}'{\sum_{m > n}}' G_{mn}^{\mbox{\small{gen}}} \cdot f_{mn}^{\mbox{\small{gen}}} \right).
\label{eq:betaoffnew}
\end{equation}

Since by construction $G_{mn}^{\mbox{\small{gen}}} \cdot f_{mn}^{\mbox{\small{gen}}}$ is a dimensionless quantity, Eq. \ref{eq:betaoffnew} implies that the first hyperpolarizability scales in the same manner as the fundamental limit:
\begin{equation}
\beta^{\mbox{off}}_{xxx} \propto \beta_{xxx}^{\mbox{max}} \propto \frac{N^{3/2}}{E_{10}^{7/2}}.
\label{eq:genscalinglaw}
\end{equation}

Alternatively, we can derive the scaling law for the first hyperpolarizability using nondimensionalization techiques. We begin by  explicitly writing Equation \ref{eq:betapairs} in the off-resonance regime as:
\begin{equation}
\beta_{xxx}^{\mbox{off}}= (-e)^{3} {\sum_{n}}'{\sum_{m > n}}'  x_{0m}x_{mn}x_{n0} \cdot F_{nm}^{\mbox{off}},
\label{eq:betaoffdf}
\end{equation}
and introduce the following dimensionless parameters:\cite{kuzyk13.01}
\begin{equation}
\xi_{nm}=\frac{x_{nm}}{\sqrt{\frac{\hbar^{2}N}{2mE_{10}}}}.
\end{equation}
This yields:
\begin{equation}
\beta_{xxx}^{\mbox{off}}= (-e)^{3} \left( \sqrt{\frac{\hbar^{2}N}{2m E_{10}}} \right)^{3} {\sum_{n}}'{\sum_{m > n}}'  \xi_{0m}\xi_{mn} \xi_{n0} \cdot F_{nm}^{\mbox{off}},
\end{equation}
or:
\begin{eqnarray}
\beta_{xxx}^{\mbox{off}} = 6 \left( \frac{e \hbar}{\sqrt{2m}} \right)^{3} \frac{N^{3/2}}{E_{10}^{7/2}} \times \\ \nonumber
{\sum_{n}}'{\sum_{m > n}}'  \xi_{0m} \xi_{mn} \xi_{n0} \sqrt{E_{n0}E_{m0}(E_{m0}-E_{n0})} \cdot \left( \frac{E_{10}}{E_{n0}} \right)^{7/2} f(\frac{E_{n0}}{E_{m0}}),
\label{eq:betaoffstep2}
\end{eqnarray}
where we have made use of the following identity:
\begin{equation}
\frac{1}{\sqrt{E_{n0}E_{m0}(E_{m0}-E_{n0})}} \cdot F_{nm}^{\mbox{off}} =-6 \cdot f(\frac{E_{n0}}{E_{m0}}) \frac{1}{E_{n0}^{7/2}}.
\end{equation}
Recalling Eq. \ref{eq:fundlim} and introducing the dimensionless parameters $e_{i}=\frac{E_{i0}}{E_{10}}$,\cite{kuzyk13.01} we can express Eq. \ref{eq:betaoffstep2} as:
\begin{equation}
\frac{\beta_{xxx}^{\mbox{off}}}{\beta_{xxx}^{\mbox{max}}} = \frac{3^{3/4}}{\sqrt{2}} {\sum_{n}}'{\sum_{m > n}}'  \xi_{0m} \xi_{mn} \xi_{n0} \sqrt{e_{n}e_{m}(e_{m}-e_{n})} \cdot f_{nm}^{\mbox{\small{gen}}}.
\label{eq:betaoffsimple}
\end{equation}
By comparing Equations \ref{eq:betaoffnew} and \ref{eq:betaoffsimple}, we conclude that $G^{\mbox{\small{gen}}}_{mn}$ must be defined as:
\begin{equation}
G^{\mbox{\small{gen}}}_{mn} = \frac{3^{3/4}}{\sqrt{2}} \xi_{0m} \xi_{mn} \xi_{n0} \sqrt{e_{n}e_{m}(e_{m}-e_{n})}.
\label{eq:gmndefin}
\end{equation}
As expected, $G^{\mbox{\small{gen}}}_{mn}$ is a dimensionless function. Furthermore, if we rewrite Equation \ref{eq:genscaling} as:
\begin{equation}
\xi_{mn} \propto \frac{1}{(e_{m}-e_{n})}.
\label{eq:genscalingdim}
\end{equation}
it becomes clear from the definition of $G^{\mbox{\small{gen}}}_{mn}$ that the energy dependence is canceled, such as (also as expected) $G^{\mbox{\small{gen}}}_{mn}$ does not depend explicitly on energies.

In conclusion, nondimensionalization techniques confirm the previous results and provide for a general expression for the $G^{\mbox{\small{gen}}}_{mn}$ function.

\subsection{The Clipped harmonic oscillator}

In order to understand better how $G_{mn}^{\mbox{\small{gen}}}$ and $f_{mn}^{\mbox{\small{gen}}}$ determine the first hyperpolarizability  let us investigate their behavior for the ``clipped harmonic oscillator'' (CHO) model, an exactly solvable model that yields $|\beta_{xxx}^{\mbox{\small{off}}}| \approx 0.57 \cdot \beta_{xxx}^{\mbox{\small{max}}}$.\cite{tripa04.01,perez04.01}

It is interesting to consider first the regular harmonic oscillator, where the potential is given by $V(x)=\frac{1}{2} m \omega^{2}$. In this case the eigenenergies are given by $E_{n}=(n+1/2)\hbar \omega$ with $n=0,1,2,\cdots$ and the only non-zero transition dipole moments are:
\begin{equation}
x_{m(m-1)}=\sqrt{\frac{\hbar}{2m\omega}} \propto \sqrt{\frac{1}{E_{m0}-E_{(m-1)0}}}.
\end{equation}
Clearly, Equation \ref{eq:genscalinglaw} is obeyed. However, all the combinations of three states that contribute to $G_{mn}^{\mbox{\small{gen}}}$ contain a null transition dipole moment such as $G_{mn}^{\mbox{\small{gen}}}=0$ for all values of $m$ and $n$. This is what we expect since due to symmetry, the simple harmonic oscillator yields a null first hyperpolarizability. 

The clipped harmonic oscillator is defined by the non-symmetric potential:
\begin{equation}
V(x)=\left\{ \begin{array}{cc}
		\infty & \mbox{for $x<0$} \\
		\frac{m\omega^{2}x^{2}}{2} & \mbox{for $x\geq 0$}.
		\end{array}
	\right.
\end{equation}

The eigenergies are given by $E_{n}=(2n+3/2)\hbar \omega$ with $n=0,1,2,\cdots$ and the transition dipole moments are given by:\cite{tripa04.01,perez04.01}
\begin{equation}
x_{ij}= \sqrt{\frac{\hbar}{\pi m \omega}} \cdot \varphi(i,j),
\label{eq:chotran}
\end{equation}
with the dimensionless function $\varphi(i,j)$ defined as:
\begin{equation}
\varphi(i,j)= 2^{-(i+j)} \frac{1}{\sqrt{(2i+1)!(2j+1)!}} \cdot \int_{0}^{\infty} H_{2i+1}(\lambda) \lambda H_{2j+1} (\lambda) d \lambda, 
\label{eq:varphi}
\end{equation}
where $H_{n}(x)$ is the $nth$ order Hermite Polynomial; and with $i=0,1,2,\cdots$ and $j=0,1,2,\cdots$.

Using that $m>n$, we confirm the predicted dependence on energy:\footnote{This follows from the fact that the factorial $(2m+1)!$ must contain the factor $2(m-n)$ since $m>n$.}
\begin{equation}
x_{mn} \propto \frac{1}{\sqrt{2(m-n)\hbar \omega}} = \frac{1}{\sqrt{(E_{m0}-E_{n0})}}.
\end{equation}
Substituting Eq. \ref{eq:chotran} into Eq. \ref{eq:gmndefin} we obtain:
\begin{equation}
G_{mn}^{\mbox{\small{gen}}}(CHO)=\frac{3^{3/4}2}{\sqrt{\pi}} \varphi(0,m) \varphi(m,n) \varphi(n,0) \cdot \sqrt{m \cdot n \cdot (m-n) }.
\label{eq:Gmncho}
\end{equation}
The energy functions are the same as for the simple harmonic oscillator and given by:
\begin{equation}
f_{nm}^{\mbox{\small{gen}}}(CHO)=\frac{1}{n^{7/2}} \cdot f(n/m).
\label{eq:Fmncho}
\end{equation}

\begin{figure}
  \centering
  \includegraphics[width=3.5in]{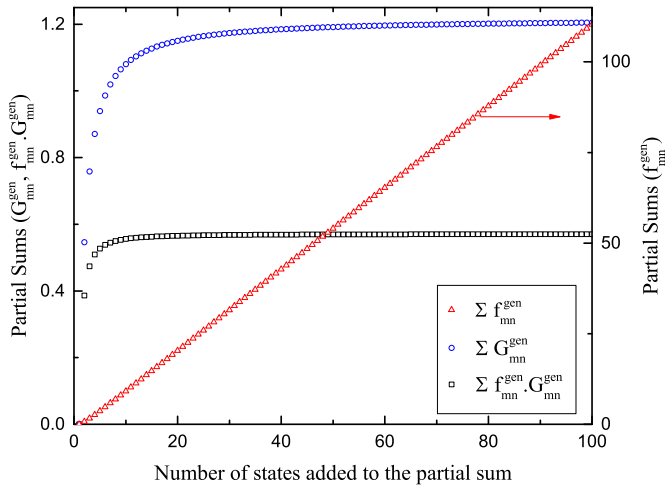}\\
 \caption{Partial sums as a function of the number of states added in the expression for the clipped harmonic oscillator (CHO): $\sum f_{mn}^{\mbox{\small{gen}}}$ (triangles), $\sum G_{mn}^{\mbox{\small{gen}}}$ (circles) and $\sum G_{mn}^{\mbox{\small{gen}}} \cdot f_{mn}^{\mbox{\small{gen}}}$ (squares).} \label{fig:Cho}
\end{figure}

The partial sums as a function of the number of added states to the sum are plotted in Figure \ref{fig:Cho}. As expected, the total sum (the product of $G \cdot f$) converges quickly to $0.57$. The convergence is fast, such as the contribution of the first three-states contribution is about 70 \% of the total sum. The partial sum of $G_{mn}^{\mbox{\small{gen}}}(CHO)$ terms converges following a similar trend to $1.2$. However, the partial sum of generalized energy functions does not converge, but increases linearly with the number of states added to the sum.\footnote{This might seem puzzling at first, since $f_{mn}^{\mbox{\tiny{gen}}} \leq 1$, but this does not guarantee convergence. For example, the infinite sum $\sum (1/n) $ does not converge.} This means that for the clipped harmonic oscillator the convergence of the total sum is due {\em only} by the convergence of the transition dipole terms, $G_{mn}^{\mbox{\small{gen}}}$. 

\subsection{Optimization and the three level ansatz}

Now let us look for potential strategies to optimize the first hyperpolarizability. 

We consider first the generalized energy functions. From the definition (Eq. \ref{eq:Fnm}) it follows that $0 \leq f_{mn}^{\mbox{\small{gen}}} \leq 1$ and that $f_{mn}^{\mbox{\small{gen}}}$ decreases in magnitude as $n$ becomes large. In the limit when $E_{m0}$ and $E_{n0}$ approach infinity, $f_{mn}^{\mbox{\small{gen}}}$ approaches zero. However, this does not imply that the partial sums of $f_{mn}^{\mbox{\small{gen}}}$ must converge, since as we shall see, when $m$ and $n$ become large there are many more contributions to the sum-over-states. 

If we assume that a specific term $f_{ij}^{\mbox{\small{gen}}}$ is optimized (i.e. $E_{i0} \gg E_{j0}$) then any term of the form $f_{jk}^{\mbox{\small{gen}}}$ has to be far from optimization (since $j>k$ implies $E_{j0} > E_{k0}$). In other words, if we try to optimize one generalized energy function term we immediately force many other terms to be far from optimization. So, when trying to optimize the first hyperpolarizability we either concentrate on optimizing a few terms, and let the other contributions to be negligible; or if we want many terms to contribute we will be forced to use energy functions that are far from optimization.

With regards to the generalized transition dipole functions $G_{mn}^{\mbox{\small{gen}}}$, we first notice that using $0 \leq f_{mn}^{\mbox{\small{gen}}} \leq 1$, we can set the partial sums of $|G_{mn}^{\mbox{\small{gen}}}|$ as an upper bound upon the magnitude of the first hyperpolarizability:
\begin{equation}
|\beta_{xxx}^{\mbox{\small{off}}}|=\beta_{xxx}^{\mbox{max}} \cdot {\sum_{n}}'{\sum_{m > n}}' |G_{mn}^{\mbox{\small{gen}}}| \cdot |f_{mn}^{\mbox{\small{gen}}}| \leq  \beta_{xxx}^{\mbox{max}} \cdot {\sum_{n}}'{\sum_{m > n}}' |G_{mn}^{\mbox{\small{gen}}}|.
\label{eq:betaIneq}
\end{equation}
This inequality holds for any number of states included in the partial sums and not only when the series converge. Indeed, we can see that this is the case for the clipped harmonic oscillator by inspecting Fig. \ref{fig:Cho}. 

In general, the values of  $G_{mn}^{\mbox{\small{gen}}}$ can be positive or negative. Since $0 \leq f_{mn}^{\small{gen}} \leq 1$, the sign of the generic term in sum over states expression (Equation \ref{eq:betaoffnew}) is determined by the sign of $G_{mn}^{\mbox{\small{gen}}}$. In most situations, the contributions of different terms to the total sum will partially cancel each other. This suggests that there must be an optimal number of contributing states where the positive effects outbalance the negative effects.

In any case, we can look at the problem of optimizing the first hyperpolarizability from a different perspective by simply counting the number of parameters that need to be manipulated in order to achieve optimization. Let us assume that the off-resonant first hyperpolarizability is optimized globally by a specific set of energies and transition moments, and that a total of $n_{tot}$ states are significantly contributing to the first hyperpolarizability (such as we can ignore the contribution of higher states). This means that the sum-over-states is represented by the contributions of $(n_{tot}-2)(n_{tot}-1)/2$ terms and each term depends on 5 parameters (3 transition dipole moments and 2 energies). Thus, the number of specific energy and transition dipole moment values that need to be finely tuned in order to achieve optimization scales as: $5(n_{tot}-2)(n_{tot}-1)/2$. Table \ref{tab:parameters} list the number of parameters that determine the first hyperpolarizability as function of the total number of states that contribute to the sum (including ground). Thus, although global optimization might be feasible computationally, such strategy will be very hard (if not impossible) to implement in the physical world if it requires the fine tuning of a large set of energies and transition dipole moments, as it would need a level of control of molecular properties that is beyond our current capabilities.

\begin{table*}
\caption{Number of parameters (energy and transition dipole moments) that determine the first hyperpolarizability as function of the total number of states that contribute to the sum (including ground), $n_{tot}$.}\label{tab:parameters}
  \centering
  \begin{tabular}{|c | c | c | c | c | c | c | c | c | c | c | c | c | c | c |}
  \hline
  Total number of contributing states ($n_{tot}$) & 3 & 4 & 5 & 6 & 7 & 8 & 9 & 10 & 11 & 12 & 13 & 14 & 15 \\
  \hline
  Number of parameters needed to determine $\beta_{xxx}^{\mbox{\tiny{off}}}$ & 5 & 15 & 30 & 50 & 75 & 105 & 140 & 180 & 225 & 275 & 330 & 390 & 455 \\
  \hline
  \end{tabular}
\end{table*}
However, we can choose to optimize the first hyperpolarizability locally, by focusing on the optimization of one of the terms in the sum. In this case, we can not guarantee that the first hyperpolarizability is reaching a global maximum, but the local maximum that we find can be achieved by tuning the smallest possible set of energies and transition dipole moments. According to table \ref{tab:parameters}, this goal is much more reasonable and easier to implement than any global optimization prescription that requires the contribution of more than three states. Since all measured compounds fall below the fundamental limit, we can focus first on designing structures that optimize the response locally, while we learn more about what is required to optimize the first hyperpolarizability globally.

As expected, when the expression for the first hyperpolarizability and the sum-rules are well represented by contributions of only three states (including ground), optimization is achieved in the same manner as predicted using the three-level ansatz: $G_{mn}^{\mbox{\small{gen}}}$ is optimized when $\xi_{n0} =\sqrt[-4]{3}$ where it reaches unity; and $f_{mn}^{\mbox{\small{gen}}}$ is optimized when the energy ratio approaches zero. In the specific case when $n=1$ and $m=2$, $G_{21}^{\mbox{\small{gen}}}$ becomes $G(X)$ and $f_{21}^{\mbox{\small{gen}}}$ becomes $f(E)$, such as we recuperate Eqs. \ref{eq:betaEX}, \ref{eq:fE} and \ref{eq:GX}. If we concentrate on optimizing the response of any other set of states, the maximum that we obtain is given by $\left( \frac{E_{10}}{E_{n0}} \right)^{7/2} \cdot \beta_{xxx}^{\mbox{\small{max}}}$.

Finally, we note that although by counting energies and transition dipole moments the number of parameters scales quadratically with the number of contributing states, these have to be over specified as they are still connected through the sum-rules. In fact, we can show that every first hyperpolarizability (that is below the fundamental limit) can be represented by two parameters, $\hat{X}$ and $\hat{E}$, defined such as the following identity is obeyed:
\begin{equation}
G(\hat{X})\cdot f(\hat{E}) = {\sum_{n}}'{\sum_{m > n}}' G_{mn}^{\mbox{\small{gen}}} \cdot f_{mn}^{\mbox{\small{gen}}} =\frac{\beta_{xxx}^{\mbox{\small{off}}}}{\beta_{xxx}^{\mbox{\small{max}}}} \equiv \beta_{int}.
\end{equation}
As long as the hyperpolarizability is below the fundamental limit we can always solve for $\hat{X}$ and $\hat{E}$. This is in agreement with independent findings that at most two parameters are important for the optimization of the first hyperpolarizability with 1-dimensional potentials.\cite{ather12.01,burke16.01}. It also confirms the validity of the quantum limits analysis for the intrepretation of experimental data.\cite{tripa04.01,tripa06.01} When all the experimental data ($\beta_{int}$, $G(X)$ and $f(E)$) is in agreement, we know that effectively three states dominate the response with $X=\hat{X}$ and $E=\hat{E}$. If it is not, we can immediately conclude that more than three states contribute to the response and use $G(X)$ and $f(E)$ as proxy functions.

\section{Conclusions}

We have shown (without approximations) that the dipole-free sum-over-states expression for the diagonal component of the first hyperpolarizability can be expressed as a sum where each term in the sum includes {\em} all the possible contributions of three specific states (including ground). This implies that in order to be consistent with the sum-rules {\em at least} three levels must significantly contribute to the response even in structures where the conjugated path is approximated to be 1-dimensional.

In systems that were well approximated as 1-dimensional governed by a time-reversal invariant Hamiltonian, the transition dipole moments have to be real. This is the case if relativistic and magnetic effects can be neglected. When the transition dipole moments are real, the expression for the first hyperpolarizability is expressed as a sum of similar terms, where each term is written as the product of three transition dipole moments ($x_{0m}x_{mn}x_{n0}$) and a energy function ($F_{nm}$).We show that tuning the energy spectrum of a molecule allows to selectively minimize or enhance the resonant response. The response is always largest in the regime of operation of the electro-optic effect. However, the best spacing for on-resonant second-harmonic generation occurs when the two energies are spaced like a two-state quantum oscillator. 

When we focus on the off-resonant response, we are able to show that, aside from the factor $E_{n0}^{-7/2}$, the energy dependence of a general term is the same as what is predicted by applying the three-level ansatz. We introduce generalized scaling laws for the transition dipole moments and prove that the first hyperpolarizability must scale in the same manner as the fundamental limit. In addition, we generalize the results derived using the three-level ansatz by expressing every contribution the the sum-over-states as a product of the fundamental limit and two dimensionless functions: $G_{mn}^{\mbox{\small{gen}}}$ and $f_{mn}^{\mbox{\small{gen}}}$. This allows to better discern how the distribution of transition dipole moments and the energy spacing affect the first hyperpolarizability. We derive this result first using the generalized scaling laws (Eq. \ref{eq:genscaling}); and using nondimensionalization techniques without invoking Eq. \ref{eq:genscaling}. Thus, even if the generalized scaling laws needed to be corrected, the results will still hold. 

We apply these principles to the clipped harmonic oscillator model and find the convergence of the first hyperpolarizability sum is due {\em only} to the convergence of the generalized transition dipole moment functions, $G_{mn}^{\mbox{\small{gen}}}$, and that  the first three-state contribution term carries $70 \%$ of the weight in the infinite sum. 

We then show than in a system with many contributing levels the generalized energy functions can not all be optimized at once. We also prove that the absolute value of the first hyperpolarizability is bounded by the set of partial sums $|G_{mn}^{\mbox{\small{gen}}}|$, and that the sign of $G_{mn}^{\mbox{\small{gen}}}$ determines the sign of every term in the first hyperpolarizability sum. As the number of states that contribute significantly to the sum-over-states increases, the chances of partial cancellation between terms increases also, so we conjecture that there must be an optimal (finite) number of states where the positive effects outbalance the negative effects. We argue that although global optimization of the first hyperpolarizability  might be possible mathematically when many states contribute to the response, the strategy will be impractical if it requires the fine tune of many molecular parameters. A more realistic approach is to optimize the response locally, by putting our efforts into the optimization on one of the three-state contributions, as prescribed by the three-level ansatz. We also confirm the validity of the fundamental limits analysis for the interpretation of experimental data.

In conclusion, we have showed that most results derived using the three-level ansatz are general and apply to molecules where more than three levels contribute to the second-order nonlinear response or/and far away from optimization. Finally, we would like to note that although the analysis presented in this paper focuses on the molecular second-order nonlinear response, the generalization to the macroscopic level is straightforward. 
 
\section{Acknowledgements}

We acknowledge Skidmore College for generously supporting this work by funding a full year sabbatical (and sabbatical enhancement) leave.
\bibliography{\bibs}

\end{document}